\documentclass[12pt]{article}
\title{  SPH \\ 
Compressible Turbulence }
\author{ J. J. Monaghan\\
 \small{School
of Mathematics Sciences}\\ \small { Monash University, Clayton 3800, Australia}\\
\small {email: joe.monaghan@sci.monash.edu.au } }
\date{\small {2 January 2002}}
\begin{document}
\maketitle
\vskip3cm
\begin{abstract}

In this paper an SPH version of the alpha turbulence model devised by Holm and his
colleagues is formulated for compressible flow with a resolution that varies in space and
time. The alpha model involves two velocity fields. One velocity field is obtained from
the momentum equation, the other by averaging this velocity field as in the version of SPH
called XSPH. The particles (fluid elements) are moved with the averaged velocity. In
analogy to the continuum alpha model we obtain a  particle Lagrangian from which the SPH
alpha equations can be derived. The system satisfies a discrete Kelvin circulation
theorem identical to that obtained with no velocity averaging. In addition, the 
energy, linear and angular momentum are conserved. We show that the continuum
equivalent of the SPH equations are identical to the continuum alpha model, and
we conjecture that they will have the same desirable features of the continuum model
including the reduction of energy in the high wave number modes even when the dissipation
is zero.  Regardless of issues concerning turbulence modelling, the SPH alpha model is a
powerful extension of the XSPH algorithm which reduces disorder at short length scales
and retains the constants of the motion. The SPH alpha model is simple to implement.

\end{abstract}

\newpage
\section{Introduction}

Many SPH simulations of astrophysical flows have regions where the SPH particles pass
through each other. In a continuum simulation this streaming is unphysical because it
implies that the velocity field is locally multi-valued. The problem can be cured by
introducing sufficient viscosity, but this has undesirable side effects, for example
excessive decay of shear. For this reason the version of SPH called XSPH was devised
(Monaghan 1989, Monaghan 1992). The idea is simple. Each particle is moved with an
smoothed velocity which is constructed by averaging the velocities of its neighbours. In
this way the counterstreaming is prevented or greatly reduced  The procedure is
non-dissipative and conserves linear and angular momentum. The smoothing has the further
advantage of reducing local disorder. Its defect is that energy is not conserved. 

The way to conserve energy, and at the same time greatly enlarge the application of SPH is
provided by the alpha model of turbulence (for early references see Holm, 1999). It is a
minimalist model of turbulence which aims to take into account the finite resolution of
any numerical simulation or experimental measurement by using an average velocity to move
the fluid, while at the same time preserving the invariants of the flow. The latter is
achieved by deriving the equations of motion from a Lagrangian which automatically
provides the stress terms which are guessed in other models of turbulence (a good modern
discussion of these models is given by Pope (2000)). The alpha equations preserve the
circulation of the fluid (when this is an invariant of the original equations) as well as
the additive integrals of energy and momentum. 

A feature of the continuum alpha model is that the presence of velocity gradient terms in
the force greatly reduces the deposit of energy in the high wave number modes. In the SPH
version the force on a  particle contains non linear terms which depend on the velocities
of neighbouring particles. We conjecture that these terms drive the energy back and forth
from the low and high wave numbers. A related example is the Fermi-Pasta-Ulam problem
(Fermi et al. 1955, Segr$\grave e$ 1965) where non linear force terms lead not to
statistical equilibrium (which would mean most energy at the high wave numbers), but to a
flow of energy back and forth amongst a few of the modes with low wave number. A similar
back and forth flow occurs in three dimensional turbulence (Piomelli et al. 1991, Jimenez
1994 , Woodward et al. 2001, Pope 2000) although this is often obscured by describing
just the nett flux which is from low to high wave numbers. In fact, at any time,
different parts of a turbulent system can have energy flow either up or down the wave
number scale.

 The advantage of the alpha model is that, if the non dissipative system already reduces
the energy in the high wave number modes, then it is possible to introduce weak
dissipation to complete the task of removing energy. Other turbulence models try
to do this by using a standard form of Navier Stokes equations with a dissipation strong
enough to remove the energy at high wave numbers. 

In this paper we derive the SPH alpha equations from a particle Lagrangian using the
continuum constant density equations as a guide. We show that the system conserves a
discrete form of the Kelvin circulation theorem in addition to the additive integrals of
energy and momentum. The equations are simple to implement and the extra computational
cost is negligible. Because the turbulence model only requires minimal dissipation, it
should be possible to reduce the normal SPH viscosity significantly in regions of the
flow away from shocks. By combining this with the idea of varying the viscosity
coefficient for each particle separately (Morris and Monaghan 1997) the viscosity could
be reduced by a factor $\sim 100$ away from shocks. 

The plan of this paper is as follows. We derive a standard set of equations for SPH from
a Lagrangian with a resolution length which is consistent with the density. The momentum
equation is found to take the same form as that obtained by Springel and Hernquist (2001)
using Lagrange multipliers to constrain the smoothing length. However, the thermal energy
equation differs from any of their thermal energy equations. We show that the equations
of motion lead to a discrete Kelvin circulation theorem in addition to the additive
integrals of energy and momentum. We then define the XSPH velocity averaging and show
that it agrees with the continuum alpha model. Guided by the alpha model we propose a
discrete SPH Lagrangian. The Lagrangian equations of motion agree in the continuum limit
with the equations of the alpha model. The conservation of circulation takes exactly the
same form as in the case when the velocities are not averaged. Finally we discuss the
dissipation and a scaling argument which suggests that the energy spectrum will fall off
more rapidly than the Kolmogorov spectrum at short length scales. The extensive numerical
work required to confirm the conjectures and the scaling argument will be presented
elsewhere. 

\section{The Comparison equations}
\setcounter{equation}{0}
\subsection{The momentum equation}
In this section we derive a standard set of SPH equations in the absence of velocity
averaging to provide a comparison with the alpha model equations. This standard set of
SPH equations will be obtained for a Lagrangian allowing fully for the variation of
resolution length $h$ with density. The equations are similar to those of Springel and
Hernquist (2001), and Bonet (1999, 2001) however the energy equation differs from those
considered by Springel and Hernquist (2001). The formulation can be easily extended to the
relativistic case using the analysis of Monaghan and Price (2001).

The Lagrangian for compressible, non dissipative flow is (Eckart 1960) 
\begin{equation}
L = \int \rho \left ( \frac{1}{2} v^2 - u(\rho,s) \right ) {\bf dr},
\end{equation}
where $u(\rho,s)$ is the thermal energy per unit mass which is a function of density
$\rho$ and entropy $s$. The SPH form of (2.1) with self gravity included is
\begin{equation}
L = \sum_b m_b \left ( \frac{1}{2} v_b^2 - u(\rho_b,s_b) + \frac{G}{2} \sum_k  \frac{
m_k m_b}{(|{\bf r}_b - {\bf r}_k|)} \right ), 
\end{equation}
where
\begin{equation}
\frac {d {\bf r}_a}{dt} = {\bf v}_a.
\end{equation}
In these equations $m_b$ is the mass, ${\bf v}_b$ the velocity, ${\bf r}_b$ the position,
$s_b$ the entropy, and $\rho_b$ the density of SPH particle $b$. In practice the
gravitational term is smoothed to remove the singularity when ${\bf r}_b = {\bf r}_k$.

In the SPH formulation the density of particle $a$ can be written 
\begin{equation}
\rho_a = \sum_b m_b W_{ab}(h_a),
\end{equation}
where $h_a$ is assumed to be a specified function of $\rho_a$ which we denote by
$H(\rho_a)$ or $H_a$. In many astrophysical calculations $H_a  \propto (1/\rho_a^{1/d})$
where the number of dimensions is $d$, but a more general function could be used. For
example, to prevent arbitarily large $h$ when $\rho$ becomes very small we
could choose
$$
H_a = \frac{A}{1+B\rho_a^{1/d}},
$$
where $A$ and $B$ are constants. Furthermore, while the usual practice is to estimate
$\rho_a$ at a given time using the value of $h_a$ from a previous time, it would be
possible and more consistent to calculate $\rho_a$ self consistently from (2.4) as
suggested recently by Bonet (2001). Equation (2.4) is a function defining the scalar
quantity $\rho_a$ and it can be solved by any standard root solving algorithm using
either the value at the previous time step as a starting value, or by estimating the
value using the rate of change of the density.

The equations of motion follow from varying the action keeping the entropy
constant. From Lagrange's equations for particle $a$  
\begin{equation}
\frac {d}{dt} \left (  \frac {\partial L}{\partial {\bf v}_a} \right ) - \frac
{\partial L}{\partial {\bf r}_a} = 0,
\end{equation}
we find
\begin{equation}
\frac {d {\bf v}_a}{dt} = -\sum_b m_b \left ( \frac {\partial u}{\partial \rho}
 \right )_s
\frac{\partial \rho_b}{\partial {\bf r}_a} - G \sum_b \frac{ m_b({\bf r}_a - {\bf
r}_b)}{|{\bf r}_a - {\bf r}_b|^3}.
\end{equation} 
 From (2.4) 
\begin{equation}
\Omega_b \frac{\partial \rho_b}{\partial {\bf r}_a} =   \sum_c m_c
\nabla_a W_{ac}(h_a) \delta_{ab} - m_a \nabla_b W_{ab}(h_b) ,
\end{equation} 
where $\nabla_a$ denotes the gradient of $W$ taken with respect to the coordinates of
particle $a$ keeping $h$ constant (we keep this convention throughout this paper), and
\begin{equation}
\Omega_b = 1 - H'_b \sum_c m_c \frac{\partial W_{bc}(h_b)}{\partial h_b}.
\end{equation}
where $H'$ denotes $\partial H_b /\partial \rho_b$. If the density variation is smooth
then $\Omega = 1 + O(h^2)$.

The first law of thermodynamics, when the entropy is constant, gives
\begin{equation}
\left( \frac {\partial u}{\partial \rho} \right )_s = \frac{P}{\rho^2},
\end{equation}
where $P$ is the pressure. The acceleration equation (2.6) with (2.7) and (2.9) then
becomes
\begin{equation}
\frac {d {\bf v}_a}{dt} = -\sum_b m_b \left ( \frac{P_a}{\Omega_a \rho_a^2} \nabla_a
W_{ab}(h_a) +\frac{P_b}{\Omega_b \rho_b^2} \nabla_a W_{ab}(h_b)\right ) - G \sum_b \frac{
m_b({\bf r}_a - {\bf r}_b)}{|{\bf r}_a - {\bf r}_b|^3}.
\end{equation}

The equation of motion (2.10) is exactly the same as the equation of motion due to
Springel and Hernquist (2001) who introduce constraints on $h$ with a Lagrange
multiplier. If the fluid is in a container the container can be represented by boundary
forces (Monaghan 1994, Monaghan and Kos 1999).

\subsection{Conservation Laws}

The symmetry of the Lagrangian leads immediately to the conservation laws. In
particular the density (and therefore P when $s$ is constant) and the gravitational
potential are independent of translations and rotations of the coordinate system. 
Accordingly, linear and angular momentum are conserved. If the system is viscous then the
viscous term can be constructed so that the linear and angular momentum are still
conserved (see $\S 2.3) $. For the constant entropy case the conservation of energy can
be shown directly by noting that
\begin{equation}
  \frac{d}{dt} \left (\sum_a {\bf v}_a \cdot \frac{\partial L}{\partial {\bf v}_a}-L
\right ) = 0.
\end{equation}

If the entropy remains constant during the motion the particle system is invariant to
other transformations. Consider, for example a set of particles each with the same mass
and entropy marking out a necklace. Imagine each particle in the loop being shifted to its
neighbour's position (in the same sense around the necklace) and given its neighbour's
velocity. Since the entropy is constant, nothing has changed, and the Lagrangian is
therefore invariant to this transformation.  A discussion of the resulting invariant
quantity has been given by Monaghan and Price (2001). We repeat the key elements of the
argument here because we intend to show (see
$\S6$) that despite moving the fluid with a smoothed velocity field the same invariant
holds.

The change in $L$ can be approximated by 
\begin{equation}
\delta L =  \sum_j \left ( \frac {\partial L}{\partial {\bf r}_j }\cdot \delta {\bf r}_j +
\frac {\partial L}{\partial {\bf v}_j } \cdot \delta {\bf v}_j \right ),
\end{equation}
where $j$ denotes the label of a particle on the loop. Because the system is invariant to
the transformation we take $\delta l =0.$ The change in position and velocity are given by
\begin{equation}
\delta {\bf r}_j = {\bf r}_{j+1} - {\bf r}_j,
\end{equation}
and
\begin{equation}
\delta {\bf v}_j = {\bf v}_{j+1} - {\bf v}_j.
\end{equation}
Using Lagrange's equations we can rewrite (2.12) in the form   
\begin{equation}
 \delta L =  \frac {d}{dt} \left (\sum_j \frac {\partial L}{\partial {\bf v}_j } \cdot
({\bf r}_{j+1} - {\bf r}_j) \right ) =  0.
\end{equation}
Recalling that the particle masses are assumed identical, we deduce that
\begin{equation}
\frac{d}{dt}  \sum_j {\bf v}_j \cdot ({\bf r}_{j+1} - {\bf r}_j)= 0,
\end{equation}
so that
\begin{equation}
C = \sum_j {\bf v}_j \cdot ({\bf r}_{j+1} - {\bf r}_j),
\end{equation}
is conserved to this approximation, for every loop. The conservation is only approximate
because the change to the Lagrangian is discrete, and only approximated by the first
order terms. However, if the particles are sufficiently close together  (2.17)
approximates the circulation theorem to arbitrary accuracy. A related argument was used
by Feynman (1957) to establish from the invariance of the wave function that circulation
should be quantised.

The system is also invariant to the particles shifting around the loop in the opposite
sense. This gives an approximation to the circulation with the opposite sign to that
above. If these two are combined (taking account of their signs so we subtract one from
the other) we get
\begin{equation}
\frac{d}{dt}  \sum_j {\bf v}_j \cdot \frac { ({\bf r}_{j+1} - {\bf r}_{j-1})}{2}= 0.
\end{equation}
which is a better approximation to the circulation of the continuous fluid. The errors in
the discrete approximation to the circulation theorem arise because of the higher order
terms in the change to the Lagrangian. These errors become less as the spatial scale of
variation of the velocity field becomes greater than the particle separation. If the
velocity is constant then
\begin{equation}
  C = \frac {1}{2} {\bf v} \cdot \sum_j ({\bf r}_{j+1} - {\bf r}_{j-1}) = 0,
\end{equation}
as expected. If the velocity is due to rigid rotation with angular
velocity ${\bf \Omega}$,  then
 \begin{equation}
  C = \frac {1}{2} {\bf \Omega} \cdot \sum_j {\bf r}_j \times ({\bf r}_{j+1} - {\bf
r}_{j-1} ).
\end{equation}
Taking the origin of the coordinate system within the loop for convenience, the area of
the triangle made by ${\bf r}_j$ and ${\bf r}_{j+1}$ is $\frac {1}{2}( {\bf r}_j \times
{\bf r}_{j+1})$. The area is counted twice so that if the loop lies in a plane with normal
$\bf n$ 
\begin{equation}
  C =  2{\bf \Omega} \cdot {\bf n} A,
\end{equation}
where $A$ is the area  within the loop of particles. The vorticity is $2{\bf \Omega}$ and
(4.18) gives the same result as would be obtained by applying Stokes theorem to the
circulation integral. Further examples are given by Monaghan and Price (2001).   

 Salmon (1988), following Bretherton (1970) also establishes the circulation law by
appealing to the invariance to particle interchange. However, because his analysis is
based on the continuum equations, it is more complicated than our derivation.

\subsection{Comments on thermal energy}
If the entropy is constant we can obtain the rate of change of thermal energy from the
rate of change of density. The rate of change of density is obtained by differentiating
(2.4) with respect to time. We find
\begin{equation}
  \frac{d \rho_a}{dt} = \frac1\Omega_a \sum_b m_b {\bf v}_{ab} \cdot \nabla_a
W_{ab}(h_a),
\end{equation}
where ${\bf v}_{ab} = ({\bf v}_a - {\bf v}_b)$. The rate of change of thermal energy when
the entropy is constant is then given by
\begin{equation}
 \frac{du_a}{dt } =  \frac{P_a}{\rho_a^2} \frac{d\rho_a}{dt}
=  \frac{P_a}{\rho_a^2 \Omega_a } \sum_b m_b {\bf v}_{ab} \cdot \nabla_a
W_{ab}(h_a), 
\end{equation}
which disagrees with all of the thermal energy equations proposed by Springel and
Hernquist (2001).

 Our thermal energy equation can also be deduced by first forming the vector dot product
of  $m_a{\bf v}_a$ and (2.10), followed by summation over $a$. We get
\begin{equation}
\frac{dE_K}{dt} = -\sum_a \sum_b m_b m_a {\bf v}_a \cdot \left (\frac{P_a}{\Omega_a
\rho_a^2}
\nabla_a W_{ab}(h_a) +\frac{P_b}{\Omega_b \rho_b^2} \nabla_a W_{ab}(h_b)  \right) -
\frac{d E_G}{dt},
\end{equation}
where $E_K$ is the kinetic energy and $E_G$ is the gravitational energy. If the labels $a$
and $b$ are interchanged in that part of the double summation involving $P_b$ we can write
(2.24) as
\begin{equation}
\frac{dE_K}{dt}+\frac{d E_G}{dt} = -\sum_a m_a \frac{P_a}{\Omega_a \rho_a^2}\sum_b m_b
{\bf v}_{ab} \cdot
\nabla_a W_{ab}(h_a) 
\end{equation}
Conservation of total energy then requires that the right hand side of (2.14) is
$-dE_{\it th}/dt$. Thus
\begin{equation}
\frac{dE_{{\it th}}}{dt} =  \sum_a m_a \frac{d u_a}{dt} = \sum_a m_a \frac{P_a}{\Omega_a
\rho_a^2}\sum_b m_b {\bf v}_{ab} \cdot \nabla_a W_{ab}(h_a) 
\end{equation}
from which (2.23) follows. 
 
In SPH calculations viscous dissipation is included by adding    
\begin{equation}
\Pi_{ab} = - \frac {\sigma_{ab} {\bf v}_{ab} \cdot {\bf r}_{ab} }
{|{\bf r}_{ab}|},
\end{equation}
to the pressure terms (Monaghan 1997). The quantity $\sigma_{ab}$ is a positive definite
parameter, invariant to the interchange of $a$ and $b$. It is related to the actual
viscosity coefficient or to an artificial viscosity coefficient dependent on the speed of
sound (see Monaghan 1997 for the particular form given in (2.27), and Monaghan 1992 for
an alternative $\Pi_{ab}$).

Accordingly, we add the term
\begin{equation}
-\frac12\sum_b m_b \Pi_{ab} \overline {\nabla_a W}_{ab},
\end{equation}
to the right hand side of (2.10) where
\begin{equation}
\overline {\nabla_a W}_{ab} = \frac12 (\nabla_a W_{ab}(h_a) + \nabla_a W_{ab}(h_b)).
\end{equation}
 Angular and linear momentum are still conserved because
the viscous force is along the line joining the centres of the SPH particles. Following
the argument leading to (2.26) we find
\begin{equation}
 \frac{du_a}{dt } =  \frac{P_a}{\rho_a^2 \Omega_a } \sum_b m_b {\bf v}_{ab} \cdot
\nabla_a W_{ab}(h_a) - \frac12 \sum_b m_b \Pi_{ab} {\bf v}_{ab} \overline {\nabla_a
W}_{ab}. 
\end{equation}

We can write $\overline {\nabla_a W}_{ab}$ in the form $ {\bf r}_{ab} F_{ab}$, where
$F_{ab} \le 0 $. It then follows that the viscous contribution to $du/dt$ is $\ge 0$. It
is also easy to show that when the fluid is viscous the entropy change is $\ge 0$.

There is another approach to the effect of dissipation which is closely analogous to that
used in the simulation of liquids at normal temperatures and pressures. In these cases
the entropy change can usually be neglected and the dominant effect of the dissipation is
to reduce the kinetic energy to zero. This is the case for water initially sloshing in a
laboratory tank. In this case we write (2.24), with the dissipation term added, in the
form
\begin{equation}
\frac {dE^*}{dt}=\frac{d}{dt} (E_k + E_G + E_{{\it th}} ) = - \frac12 \sum_b m_b \Pi_{ab}
{\bf v}_{ab}
\overline {\nabla_a W}_{ab}, 
\end{equation}
where only the pressure terms are included in the rate of change of thermal energy.
Because the right hand side of (2.31)is negative definite the SPH equations
guarantee that dissipation will cause $E^*$ to decrease as it should. For example, an
oscillating polytrope simulated keeping the entropy of every particle constant, will
evolve to a state where $E^*$ is a minimum. When it reaches this state $E_K$ will be zero.

\section{Averaging the velocity field}
 \setcounter{equation}{0}
In the standard SPH calculations the the position of particle $a$ is found by
integrating 
\begin{equation}
\frac { d {\bf r}_a}{dt} = {\bf v}_a .
\end{equation}
However, it is possible for particles to stream through each other unless the viscosity
is large. For that reason the XSPH version of SPH was suggested (Monaghan 1989) where the
particles were moved with with a smoothed velocity, but the acceleration equation was
unchanged. The smoothed velocity ${\bf \widehat v}_a $ is defined by an average over the
velocities of neighbouring particles according to
\begin{equation}
{\bf \widehat v}_a = {\bf v}_a +  \epsilon \sum_b \frac {m_b}{\bar
\rho_{ab}} ( {\bf v}_b - {\bf v}_a) K(a,b),
\end{equation}
where $\epsilon$ is a parameter that is typically chosen to be 0.5, $\bar \rho_{ab}
$ denotes an average density, for example
\begin{equation}
\frac {1}{\overline {\rho}_{ab}} = \frac {1}{2} \left ( \frac {1}{\rho_a}+\frac
{1}{\rho_a}
\right),
\end{equation}
and $K(a,b) = K(b,a)$ is a smoothing kernel. The kernel $K(a,b)$ does not have to be the
same as the kernel used in other SPH equations. In this paper we assume the kernel has the
form 
\begin{equation}
K(a,b) = \frac12 \left ( W_{ab}(h_a) + W_{ab}(h_b) \right ).
\end{equation}

The SPH particles are moved according to
\begin{equation}
\frac{d{\bf r}_a}{dt} = \widehat {\bf v}_a.
\end{equation}

The smoothed velocity has a number of useful properties. First we note that
\begin{equation}
   \sum_a m_a {\bf v}_a = \sum_a m_a {\bf \widehat v}_a,
\end{equation}
which shows that the linear momentum is the same whether computed with $\widehat {\bf v}$
or ${\bf v}$. Second the centre of mass moves with the ${\bf v}$ because
\begin{equation}
\frac{d}{dt} \sum_a m_a {\bf r}_a = \sum_a m_a {\bf v}_a.
\end{equation}
Third, the angular momentum is unaffected because
\begin{equation}
\sum_a m_a {\bf r}_a \times \frac{d{\bf v}_a}{dt} = \frac{d}{dt} \sum_a m_a {\bf r}_a
\times {\bf v}_a.
\end{equation}

 Finally we note that the motion can be exactly reversed by reversing ${\bf v}$ and
$\widehat {\bf v}$.

 The continuum limit of (3.2) is
\begin{equation}
\widehat {\bf v}_a = {\bf v }_a + \frac{\epsilon}{2 \rho_a} \int (\rho({\bf r}) + \rho_a) 
 ({\bf v({\bf r'})} - {\bf v({\bf r}_a)} ) K({\bf r'},{\bf r}_a) {\bf dr'},
\end{equation}
where
\begin{equation}
K({\bf r'},{\bf r_a}) = \frac12 ( W({\bf r} - {\bf r}_a,h_a) +  W({\bf r} - {\bf
r}_a,h(\rho({\bf r'}))
\end{equation}
 Taylor expansion about ${\bf r}_a$ followed by integration gives the approximation
\begin{equation}
\widehat v^j_a = v^j_a + \frac{ \alpha^2}{ \rho_a} \nabla \cdot (\rho v^j) + \frac12
(\nabla \rho \cdot \nabla v^j )\frac{\partial H_a}{\partial \rho_a}\frac{\partial
\alpha^2}{\partial h_a},
\end{equation}
where
\begin{equation}
\alpha^2 =\frac{\epsilon}{2} \int (x' -x_a)^2 W({\bf r'} - {\bf r}_a,h_a) {\bf dr'},
\end{equation}
defines the $\alpha$ in the alpha model although, in our analysis, $\alpha$ may be
different for each particle and, because $h_a$ can vary with time, so can $\alpha$. If
$\rho$ is constant (3.11) reduces to 
\begin{equation}
\widehat v^j_a = v^j_a + \alpha^2 (\nabla^2 v^j)_a,
\end{equation} 
as in the alpha model.

For typical SPH kernels $\alpha^2 \propto h^2$. Accordingly,
provided the scale of variation of the velocity field is much greater than $h$, the
approximation (3.11) will be satisfactory. However, our original XSPH smoothing (3.2)
applies regardless of the scale of the velocity field variation. In particular we expect
that it can be used even when shocks occur.

The results of this section show that the XSPH smoothing has a number of attractive
features. It prevents SPH particles passing through each other, it retains the
conservation of linear and angular momentum, and it agrees with the alpha model when the
density is constant. We now show that the equations of motion can be obtained from
a Lagrangian which is a function  ${\bf r}$  and  $\widehat {\bf v}$.

\section{Lagrange's equations for XSPH}
 \setcounter{equation}{0}

The analysis of Holm (1999) suggests that the Lagrangian will take the form
\begin{equation}
L = \sum_b m_b \left ( \frac{1}{2} {\bf \widehat v }_b \cdot {\bf v}_b -u(\rho_b,s_b)
\right ),
\end{equation}
where the canonical coordinates for particle $b$ are ${\bf r}_b$ with rate of change
${\bf \widehat v}_b$. In the interests of simplicity we have not included gravitational
forces. These can be easily included in the final equations.

 The kinetic energy part of this Lagrangian can be rewritten by inverting
(3.2) to give
\begin{equation}
{\bf  v}_a = {\bf \widehat v}_a -  \epsilon \sum_b \frac {m_b}{\bar
\rho_{ab}} ( {\bf \widehat v}_b - {\bf \widehat v}_a) W_{ab},
\end{equation}
where the errors in the inversion are of $O(h^4)$. We then get
\begin{equation}
\frac{1}{2} \sum_b m_b{\bf \widehat v}_b \cdot {\bf v}_b = \frac{1}{2} \sum_b
m_b{\bf \widehat v}_b \cdot {\bf \widehat v}_b + \frac {\epsilon}{4} \sum_k \sum_b \frac
{m_b m_k}{\bar \rho_{bk}} ( {\bf \widehat v}_b - {\bf \widehat v}_k)^2 W_{kb}
\end{equation}
which is positive definite.

If the sums over $k$ are approximated by integrals, and we assume the density is
constant, Taylor series expansion of the velocity difference in the
integrand, followed by integration, shows that 
\begin{equation}
\sum_k \frac { m_k}{\bar \rho_{ab}} ( {\bf \widehat v}_b - {\bf \widehat v}_a)^2 W_{kb}
= (\nabla \widehat v_b^i \cdot \nabla \widehat v_b^i) \int (x_b - x)^2 
W({\bf r}_b - {\bf r}^\prime) d \tau^\prime,
\end{equation}
where $\widehat v_a^i$ denotes the $i$ th component of $\widehat {\bf v}_a$, and there is
an implied summation over $i$. Substituting this result into (4.3) gives
\begin{equation}
\frac{1}{2} \sum_b m_b({\widehat v}_b^2 + \alpha^2\nabla \widehat v_b^j \cdot \nabla.
\widehat v_b^j ).
\end{equation}
In the absence of dissipation this quantity is an invariant for an incompressible
fluid (Holm 1999) and is the alpha equivalent of the kinetic energy. Note particularly
the $\nabla v$ terms which play a key role in the distribution of energy. The SPH
equivalent is the sum involving the square of the velocity differences. This sum only
involves neighbour particles, and is small if they have similar velocities. It becomes
large only when the velocities of the neighbouring particles differ in sign or magnitude
or both. This occurs when there is substantial change in $v$ on the scale of $h$. Such a
change is equivalent to the existence of high wave number modes. 
 
 Returning to Lagrange's equations the canonical momentum for particle $a$ is
\begin{equation}
\frac {\partial L}{\partial {\bf \widehat v}_a } = m_a {\bf \widehat v}_a + \frac
{\epsilon}{2} \sum_k \sum_b \frac {m_b m_k}{\bar \rho_{bk}} ({\bf \widehat v}_b - {\bf
\widehat v}_k )( \delta_{ba} - \delta_{ka}) W_{kb},
\end{equation}
which reduces to 
\begin{eqnarray}
\frac {\partial L}{\partial {\bf \widehat v}_a } &=& m_a {\bf \widehat v}_a - 
\epsilon \sum_k  \frac {m_a m_k}{\bar \rho_{ak}} ({\bf \widehat v}_k - {\bf
\widehat v}_a ) W_{ka},\\
& = & m_a {\bf v}_a 
\end{eqnarray}
so that the canonical momentum of the Lagrangian defined with ${\bf \widehat v}$ is the
normal momentum constructed with ${\bf v}$. This dual relation between ${\bf v}$ and
$\widehat {\bf v}$ reminds us of the relationship between Eulerian and Lagrangian
velocities noted by Holm (1999).

The remaining term required for the equation of motion of particle $a$ is $\partial L/
\partial {\bf r}_a $. The details of this calculation are given in the Appendix 1. We find
\begin{equation}
\frac{d {\bf v}_a}{dt}= - \sum_b m_b \left ( \left ( \frac{P_a}{\Omega_a \rho_a^2} -
\frac{\epsilon}{4}  \left ( \frac{v_{ab}^2}{\bar \rho_{ab}} - \frac{\zeta_a}{\Omega_a
\rho_a^2} + \frac{2\nu_a H'_a}{\rho_a \Omega_a} \right ) \right) \nabla_a W_{ab}(h_a)  +
(a \rightarrow b)  \right ),
\end{equation}
where $(a \rightarrow b)$ denotes terms that are identical to the previous but $a$ and
$b$ are interchanged except in $\nabla_a$,  and the other quantities, where not previously
defined, have the following meaning
$$
v_{ab}^2 = ({\bf v}_a - {\bf v}_b)^2,
$$
$$
\zeta_a = \frac12\sum_k m_k v_{ak}^2 ( W_{ak}(h_a) +  W_{ak}(h_k) ),
$$
and
$$
\nu_a  = \sum_k m_k \frac {v_{ak}^2 }{\overline {\rho}_{ak}}\frac{\partial
W_{ak}(h_a)}{\partial h_a}.
$$

The equations required for the SPH alpha model are the acceleration equation (5.9), and
the density equation either in the summation form (4.2) or the continuity equation 
\begin{equation}
\frac{d \rho_a}{dt} = \frac{1}{\Omega_a}\sum_b m_b {\bf \widehat v}_{ab}  \cdot \nabla_a
W_{ab},
\end{equation}
together with $d{\bf r}_a/dt = \widehat {\bf v}_a$. The derivative following the motion is
given by 
\begin{equation}
 \frac {d }{dt} = \frac {\partial}{\partial t} + {\bf \widehat v} \cdot \nabla.
\end{equation}

These beautifully simple equations comprise the SPH alpha model. They contain all the desirable
features of Holm's (1999) incompressible flow equations but generalised to compressible
flow.  If the fluid is self gravitating then the gravity terms can be added as in $\S2$.
These new equations differ from those in $\S2$, for example (2.10), because the velocity
averaging leads to velocity dependent terms in the force and the particles are moved with
the smoothed velocity. Otherwise the equations are very similar.  

In order to compare our results with that of the incompressible alpha model we assume $h$
and $\rho $ are constant. In this case $\Omega = 1$ and the term involving $\nu$ vanishes
because $\nabla h$ is zero. The equation of motion becomes
\begin{equation}
\frac{d {\bf v}_a}{dt}= - \sum_b m_b \left ( \left( \frac{P_a}{ \rho_a^2} -
\frac{\epsilon}{4} \left ( \frac{v_{ab}^2}{\bar \rho_{ab}} - \frac{\zeta_a}{\rho_a^2}
 \right) \right) \nabla_a W_{ab}(h_a) + (a \rightarrow b) \right ).
\end{equation} 

 If the sums in the SPH equations are converted to integrals the
dominant terms give the continuum alpha model (for the details see Appendix 2). In
particular, we recover Holm's acceleration equation (Holm 1999, equation (143)) 
\begin{equation}
 \frac {\partial v^j}{\partial t} + ({\bf \widehat v} \cdot \nabla) v^j + v^\ell \frac
{\partial \widehat v^\ell}{\partial x^j} = - \frac{1}{\rho} \frac{\partial}{\partial x^j} 
\left (P - \frac{1}{2} {\bf \widehat v}\cdot {\bf
\widehat v} - \frac {\alpha^2}{2} \frac {\partial \widehat v^k}{\partial x^\ell}\frac
{\partial \widehat v^k}{\partial x^\ell} \right ).
\end{equation}

We have therefore achieved our aim of constructing a Lagrangian based turbulence model
which reduces to the alpha model in the continuum limit. In the following sections we will
investigate some of the properties of the SPH model.

\section{Conservation laws}

\setcounter{equation}{0}

In the absence of boundaries and external forces the Lagrangian is invariant to
translations and rotations. Linear and angular momentum are therefore conserved. Because
the Lagrangian has no explicit time dependence there is an invariant
\begin{eqnarray}
E &=& \sum_a  {\bf \widehat v}_a \cdot \frac {\partial L }{\partial {\bf \widehat v}_a} -
L,\\
  & = & \sum_a m_a \left ( \frac {1}{2} {\bf \widehat v}_a \cdot {\bf v}_a + u(\rho,s)
\right ).
\end{eqnarray} 
which can be appropriately called the energy. Using (5.3) we can write
\begin{equation}
E =  \sum_b m_b \left ( \frac{1}{2}{\bf \widehat v}_b \cdot {\bf \widehat v}_b + \frac
{\epsilon}{4} \sum_k  \frac { m_k}{\bar
\rho_{bk}} ( {\bf \widehat v}_b - {\bf \widehat v}_k)^2 W_{kb} + u(\rho_b,s_b) \right ).
\end{equation}

If the entropy is constant then the system is also invariant to the necklace
transformation considered earlier ($\S 2.2$. In the present case the particles, assumed to
have the same mass and entropy, are shifted to the adjacent position on the loop and
given the
${\widehat {\bf v}}$ of the new position.  

The change in $L$ can be written
\begin{equation}
\delta L =  \sum_j \left ( \frac {\partial L}{\partial {\bf r}_j } \cdot \delta {\bf r}_j
+
\frac {\partial L}{\partial { \bf \widehat v}_j }\cdot \delta { \bf \widehat v}_j \right
),
\end{equation}
where $j$ denotes the label of a particle on the loop. The change in position and
velocity are given by
\begin{equation}
\delta {\bf r}_j = {\bf r}_{j+1} - {\bf r}_j,
\end{equation}
and
\begin{equation}
\delta { \bf \widehat v}_j = {\bf \widehat v}_{j+1} - {\bf \widehat v}_j.
\end{equation}
Using Lagrange's equations we can rewrite (5.4) in the form   
\begin{equation}
 \frac{d}{dt} \left (\sum_j  \frac {\partial L}{\partial \widehat {\bf v}_j} \cdot ({\bf
r}_{j+1} - {\bf r}_j) \right )
 = 0,
\end{equation}
and recalling that the particle masses are identical, we deduce
\begin{equation}
\frac{d}{dt}  \sum_j {\bf v}_j \cdot ({\bf r}_{j+1} - {\bf r}_j)= 0.
\end{equation}
which is identical to the discrete circulation theorem (2.16). The reader will note that
this result depends on the fact that the canonical momentum of particle $j$ is $m_j {\bf
v}_j$.

\section{ Is there a steady state ?}

The SPH alpha model, in the absence of dissipation, is a conservative mechanical system
derived from a Lagrangian. If the system is at rest, for example in a uniform periodic
box, then small disturbances would propagate as sound waves. In the linear approximation
the waves are not coupled so that energy put into a mode stays there. If the motion
becomes non linear the waves are coupled and any is shared amongst the modes. If
equilibrium statistical mechanics was valid for this system each mode would have the
same energy, and the distribution of energy between wave number $k$ and $k+ dk$ would be
proportional to $k^2$ in three dimensions. However, it is known from other cases, for
example the Fermi-Past-Ulam study of coupled non linear oscillators (Fermi et. al 1955,
Segr$\grave e$), that a non linear non-dissipative system may not evolve to statistical
equilibrium, but instead moves energy back and forth between a relatively small number of
modes. 

In the present case we have non-linear coupling through both the velocity terms in the
equation of motion, and the density variations with velocity. The results of Chen et. al
(1999) for the continuum alpha model suggest that in wave number space the non-linear
terms act to turn the spectrum down for large wave numbers. It is as if velocity terms in
the equation of motion acted as a potential preventing a piling up of energy at the high
wave numbers. This suggests that the non-dissipative model transfers
energy back and forth between high and low wave numbers. 

The importance of this for turbulence is the following. It is known that a direct
numerical simulation without dissipation leads to a piling up of energy in the high wave
number modes. If dissipation is included, but it is very weak at the resolution of the
numerical simulation, then the piling up will still take place, though eventually, if
there is no input of energy it will decay. If there is an input of energy at the low wave
numbers, and it is sufficiently large, then the piling up will continue because it is
arriving faster than it can leave. The idea of sub-grid models is to prevent this piling
up by including extra dissipation. The alpha model has the ability, without dissipation,
to reduce the piling up of energy at high wave numbers. For this to work it is necessary
for the system to transfer energy back and forth between high and low wave numbers. As
noted earlier, in three dimensional turbulence there is transfer both ways (Piomelli et
al. 1991, Jimenez 1994, Pope 2000, Woodward et al. 2001) and, at any time, parts of the
domain may have transfer up, and parts may have transfer down. The downward transfer is
the residual between two much larger oppositely directed fluxes. The alpha model
appears to mimic these features. How this occurs requires a numerical investigation of the
non-dissipative system and this will be given elsewhere. 

If energy is put into the system then it is still necessary to introduce
dissipation but it can be very much weaker than the dissipation used in the usual sub-grid
models. 

\section{Dissipation}
\setcounter{equation}{0}

The previous analysis concerns the non-dissipative fluid. If it is correct that the
non-dissipative SPH model transfers energy between low and high wave numbers to prevent
the piling up of energy at high wave numbers, then it should be possible to
add a small amount of dissipation to remove energy at the resolution scale ($\sim h$ in
the SPH calculation) so that, with an energy input at low wave numbers, the system
reproduces the Kolmogorov spectrum over all but the high wave number modes.

The dissipation term for SPH is based on $\Pi_{ab}$ (see (2.27)). In the present case
where the particles are moved with $\widehat {\bf v}$ we replace the velocity ${\bf v}$ in
$\Pi_{ab}$ with $\widehat {\bf v}$, and substitute the term in the right
hand side of (4.9) to obtain the dissipative equations. For convenience we define
$$
L_K  = \sum_a m_a \frac12 \widehat {\bf v}_a \cdot {\bf v}_a,
$$
then (4.9) can be written
\begin{equation}
\frac {d}{dt} \left (\frac {\partial L_K}{\partial {\bf \widehat v}_a}\right ) - \frac
{\partial L_K}{\partial {\bf r}_a} + m_a \sum_b m_b \left ( \frac{P_a}{\Omega_a \rho_a^2}
\nabla_a W_{ab}(h_a) + ( a\rightarrow b) \right ) = - m_a \sum_b m_b \Pi_{ab}
\overline {\nabla_a W}_{ab}.
  \end{equation}
We can deduce the thermal energy equation by requiring the total energy to be
constant. Using the same arguments as in $\S 2.3$, but now taking the dot product of the
acceleration equation with $\widehat {\bf v}_a$, we get the following equation for the
thermal energy of particle

\begin{equation}
 \frac{du_a}{dt } = \frac{P_a}{\rho_a^2 \Omega_a } \sum_b m_b \widehat {\bf v}_{ab}
\cdot \nabla_a W_{ab}(h_a) - \frac12 \sum_b m_b \Pi_{ab} \widehat {\bf v}_{ab} \cdot
\overline {\nabla_a W}_{ab}. 
\end{equation}
The term involving the dissipation is positive definite so that the SPH dissipation
increases the thermal energy as it should. It is easy to deduce that the dissipation also
guarantees that the entropy increases for each particle.

In the same way $E^*$, the quasi energy appropriate for systems computed assuming
the entropy of each particle is constant (see
$\S 2.3$), changes with time according to

\begin{equation}
\frac{dE^*}{dt} = \sum_a \sum_b m_a m_b \Pi_{ab} {\bf \widehat v}_a \cdot \overline
{\nabla_a W}_{ab},
\end{equation}
so that, as in the case without velocity averaging, the system will evolve to a state
with $E^*$ a minimum.

 In order to establish these results it has been necessary to use the smoothed velocity
$\widehat {\bf v}$ instead of ${\bf v}$ in the viscosity term $\Pi_{ab}$. We can
therefore expect that the SPH viscosity will be smaller than usual in those parts of the
flow where the velocity is smooth. In SPH calculations the viscosity term $\Pi$ has a
coefficient (also denoted by $\alpha$) which can be assigned a value $\sim 1$. If
desired, each SPH particle can have its own $\alpha$ which can then be determined by the
dynamics (Morris and Monaghan 1997) in such a way that its value is $\sim 0.1$ away from
shocks. If the smoothed velocity is used in $\Pi_{ab}$, and the coefficient is varied as
described by Morris and Monaghan (1997), then the coefficient away from shocks can be
expected to be reduced and values of $\sim 0.01$ appear possible. Preliminary
investigations of polytrope oscillations and Kelvin-Helmholtz instabilities show that
this conjecture is true in these cases. A full investigation of these effects will be
given elsewhere.

\section{Scaling and Turbulence}

The previous sections have laid out the formalism of the SPH alpha model of turbulence.
If there is no dissipation we have argued that there is a back and forth flow of energy
between low and high wave numbers. If there is an input of energy at low wave numbers,
and the system has a small dissipation of the type considered above, then we conjecture
that the SPH system will reproduce the Kolmogorov spectrum for the low wave number, energy
containing modes. Support for this conjecture is provided by the continuum calculations
of Chen et al. (1999).

Some features of this process can be guessed by  scaling arguments (Kolmogorov and
Oboukov as described by Landau and Lifshitz (1993) which assume (a) that the details of
the dissipation are irrelevant to the low wave number modes and (b) the rate of loss of
energy through the various length scales is constant. 

We start with the energy

\begin{equation}
E =  \sum_b m_b \left ( \frac{1}{2}{\bf \widehat v}_b \cdot {\bf \widehat v}_b + \frac
{\epsilon}{4} \sum_k  \frac { m_k}{\bar
\rho_{bk}} ( {\bf \widehat v}_b - {\bf \widehat v}_k)^2 \overline {W}_{kb} + u(\rho_b,s_b)
\right ).
\end{equation}
and assume there is a specified rate of loss of energy from the large scale motions
$\mathcal{ E}$ and this cascades through the various length scales until it is dissipated
by either by our weak SPH viscosity or by molecular viscosity.

The rate of change of energy per unit mass $\mathcal{E}$ determined by the first term in
$E$ for velocity variations with characteristic length $\ell$ is 
\begin{equation}
\mathcal{E} =  \frac {v_\ell^3}{\ell} \left (1 + \frac {\alpha^2}{\ell^2},
\right )
\end{equation}
where the term involving $\alpha^2$ can be estimated from the continuum limit of the
energy which introduces the term $\alpha^2 \nabla v^j \cdot \nabla v^j$ (see (4.5)).
Because $\alpha$ varies with the resolution length the following scaling formula should be
interpreted in terms of the local conditions of the gas. 

The characteristic velocity is
\begin{equation}
v_\ell= \left ( \frac {\ell^3 \mathcal{E}}{\ell^2 + \alpha^2} \right )^{\frac{1}{3}}.
\end{equation}
If $\ell \gg \alpha$ then this reduces to the usual Kolmogorov velocity scaling $v_\ell =
\ell^{1/3} \mathcal{E}^{1/3}$.  If $\ell \le \alpha$
\begin{equation}
v_\ell = \ell \left ( \frac { \mathcal{E} }{\alpha^2} \right)^{\frac{1}{3} },
\end{equation} 
The velocity therefore reduces more rapidly with $\ell$ in the smoothing region. The
turnover time is given by 
\begin{equation}
\tau = \frac {\ell}{v_\ell} = \left ( \frac {\ell^2 + \alpha^2}{\mathcal{E} }\right
)^{\frac{1}{3} }.
\end{equation}
which is approximately constant when $\ell \ll \alpha$. This shows that all length scales
have essentially the same turnover time when the smoothing dominates. These effects on the
velocity scale and the turnover time are not due to dissipation. They occur because of
the quadratic velocity terms in the acceleration equation.

If the energy per unit mass in the wave number range $k$ to $k+dk$ is $E(k)dk$ then
$E(k)$ has the dimensions of $(length)^3 /(time)^2$.  We can relate this to the
energy dissipation and the wave number. For large length scales we expect that $E(k)$
will vary as $\mathcal{E}^r k^n$. We then find  
\begin{equation}
 E(k) = \frac {\mathcal{E}^{\frac{2}{3} }}{ k^{\frac{5}{3}}}.
\end{equation}
For the short length scales we relate $E(k)$ to that part of $\mathcal{E}$ which
depends on the characteristic velocity and length scales. Referring to (8.5) this is 
\begin{equation}
\eta = \frac{v_\ell^3}{\ell^3}.
\end{equation}  
We then find
\begin{equation}
E(k) = \frac {\eta^{\frac{2}{3}}}{k^3}. 
\end{equation}
The energy spectrum therefore steepens for small length scales. These approximate
results have been confirmed for the continuum case by numerical experiments (Chen et al.
1999) but a similar extensive study is required to confirm these arguments for the SPH
alpha model.

\section{ Conclusions}

The results of this paper can be summarised as follows:
\begin{enumerate}

\item An averaged or smoothed velocity can be constructed for each particle such that, if
each particle moves with this velocity, the linear and angular momentum are unchanged from
the values with the unsmoothed velocity. This part of our analysis is the same as in the
XSPH algorithm.

\item The smoothing of the velocity agrees with that used in the alpha model of
turbulence in the continuum limit with the density constant.

\item A particle Lagrangian can be constructed using as canonical variables the smoothed
velocity $\widehat {\bf v}$ and ${\bf r}$. This Lagrangian has the same structure as the
continuum Lagrangian in the alpha model.

\item The SPH Lagrangian leads to equations which conserve energy, linear and angular
momentum and satisfy the same discrete Kelvin circulation theorem as for SPH with no
smoothing of the velocity.

\item In the continuum limit the new SPH equations agree with the equations of the alpha
turbulence model when the density is constant.
\end{enumerate}

In addition we conjecture that:
\begin{enumerate}

\item  The non dissipative case should prevent energy piling up at the highest wave
numbers allowed by the resolution and it does so by transferring energy back and forth
between low and high wave numbers.  This conjecture is supported by scaling arguments, by
the fact that this occurs in laboratory turbulence, and by analogies to dynamical systems
like the Fermi-Past-Ulam problem.

\item  Weak dissipation should be adequate to reproduce the Kolmogorov spectrum for
isotropic homogeneous turbulence. This conjecture is supported by the calculations for
the continuum case by Chen et al. (1999).

\item  The artificial viscosity used in SPH calculations can be very much smaller than
that currently used in smooth parts of the flow provided each particle has its own
viscosity coefficient. This conjecture is supported by preliminary calculations of the
oscillations of a polytrope and the growth of Kelvin Helmholtz instabilities. 

\end{enumerate}
 
This paper has concentrated on the dynamics of the SPH system with and without viscous
dissipation. We have had in mind those turbulent processes which arise from mechanical
sources such as Kelvin Helmhotz instabilities in jets. We have not considered the more
complex case of turbulence produced by thermal effects as in turbulent convection. In
particular we have not considered how the smoothing of the velocity field affects the
transport of heat. The formulation of the alpha model gives us two velocity fields and
from these it appears possible to model the increased diffusivity expected from
turbulence. How this should be done, however, is not clear. 

 Another important issue in turbulence dynamics is mixing (Warhaft 2000). The problem is
often counter-intuitive, with many seemingly laminar flows leading to significant mixing
(Aref et al. 1989). The natural approach to mixing is the Lagrangian formulation and it is
possible that the SPH formulation provides an efficient practical means of determining
the extent of mixing both in turbulent and in nearly laminar flow. This is another area of
research which is opened up by the theoretical developments in this paper. 

\section { Acknowledgment}
Numerous, pleasant and thought provoking interactions with Darryl Holm have greatly
improved this paper. 

\section{ References}
\begin{enumerate}
\item Aref, H., Jones, S.W., Mofina, S., and Zadwadzki, I. {\em Physica D}, {\bf 37},
423, (1989)
\item Bonet, J., and Lok, T-S,L. {\em Comp. Meth. App. Mech. Engrg}{\bf 180}, 97, (1999)
\item Bonet, J. \emph{Meshless Methods}, Bonn (2001).
\item Bretherton, F. {\em J. Fluid Mech.}, {44}, 19, (1970).
\item Chen, S., Holm, D. D., Margolin, L. G., and Zhang, R. {\em Physica D}, {\bf 133}, 66
, (1999)
\item Eckart, C. {\em Phys. Fluids}, {\bf 3}, 421, (1960)
\item Fermi,E., Pasta, J., and Ulam, S. {\em Studies of Non Linear Problems. LA-1940}
(1955)
\item Feynman, R. {\em Prog. Low Temp. Phys.}, {\bf 1}, 17, (1957)
\item Ghosal, S. {\em AIAA J.}, {\bf 37}, 425, (1999).  
\item Holm, D. D. {\em Physica D},{\bf 133}, 215, (1999)
\item Jim$\grave e$nez, J. {\em Fluid Physics. Lecture Notes on Summer Schools.} Ed. M.G.
Verlarde and C.I. Christov.  Publ. World Scientific. (1995). 
\item Landau, L.D., and Lifshitz, E.M. {\em Fluid Mechanics Vol 6, Course of Theoretical
Physics}. (1993). Publ. Pergamon.
\item Monaghan, J.J. {\em Jour. Computat. Phys.}, {\bf 64}, 2, (1989)
\item Monaghan, J.J. {\em Ann. Rev. Astron. Ap.}, {\bf 30}, 543, (1992)
\item Monaghan, J.J.{\em Jour. Computat. Phys.}, {\bf 110}, 399, (1994)
\item Monaghan, J.J., and Kos, A., {\em J. Waterways, Ports, Coastal and Ocean Eng.},
{125}, 145, (1999)
\item Monaghan, J. J., and Price, D.L. {\em Mon. Not. Roy. Astr. Soc.}{\bf 328}, 381,
(2001)
\item Morris, J.P., and Monaghan, J.J. {\em J. Computat. Phys.},{\bf 136}, 41, (1997).
\item Piomelli, U., Cabot, W.H., Moin, P., and Lee, S. {\em Phys. Fluids A }{\bf 3},1766,
(1991).
\item Pope, S.B. \emph {Turbulent Flows}. Publ. Cambridge University Press. (2000).
\item Woodward, P., Porter. D. H., Sytine, I., Anderson, S. E., Mirin, A. A., Curtis, B.
C., Cohen, R. H., Dannevik, W. P., Dimits, A. M., Eliason, D. E., Winkler, K-H., and
Hodson S. W. \emph { Computational Fluid Dynamics}, 4th UNAM Supercomputing Conference.
Ed. E. Ramos and G. Cisneros. Publ. World Scientific.  (2001).
\item Salmon, R. {\em Ann. Rev. Fluid Mech. }, {\bf 20}, 225, (1988)
\item  E.Segr$\grave e$ Ed. \emph{ Enrico Fermi Collected Papers. Vol II}. page 266.
(1965). Univ. of Chicago Press.
\item Springel, V., and Hernquist, L. {\em Mon. Not. Roy. Astr. Soc} (2001).
\item Warhaft, Z.{\em Ann. Rev. Fluid Mech. }, {\bf 32}, 203, (2000).
\end{enumerate}

\appendix 
 \section {The $\partial W/\partial h $ terms}
\setcounter{equation}{0}

The contribution to the Lagrangian from the thermal energy gives the usual terms
involving the pressure. The velocity terms in the Lagrangian are
\begin{equation}
L_v = \frac{\epsilon}{4} \sum_b \sum_k m_b m_k \frac{v_{bk}^2}{\bar \rho_{bk} } \overline
{W}_{bk},
\end{equation}
where
\begin{equation}
\overline {W}_{bk} = \frac12( W_{bk}(h_k) + W_{bk}(h_b)).
\end{equation}
By a change of label it is easy to show that $\overline {\rho}_{bk}$ can be replaced by
$\rho_b$ in (A.1). Then
\begin{equation}
\frac{\partial L_v}{\partial {\bf r}_a } 
        = \frac{\epsilon}{4} \sum_b \sum_k m_b m_k v_{bk}^2
\left ( \overline
{W}_{bk} \frac{\partial}{\partial {\bf r}_a } \left ( \frac {1}{\rho_b} \right )  +
\frac{1}{\rho_b} \frac {\partial \overline {W}_{bk} }{\partial {\bf r}_a }  \right ).
\end{equation}
Making use of the earlier expression for the density gradient (2.7), and some
straightforward simplification, the first term in (A.3) can be written 
\begin{equation}
   - \frac{m_a \epsilon}{4} \sum_b \left ( \frac{\zeta_a}{\rho_a^2 \Omega_a}
\nabla_a W_{ab}(h_a) + \frac{\zeta_b}{\rho_b^2 \Omega_b}
\nabla_a W_{ab}(h_a) \right),
\end{equation}
where 
\begin{equation}
\zeta_a = \sum_k m_k v_{ak}^2 \overline {W}_{ak}.
\end{equation}

The second term in (A.3) is more complicated. We first consider the contribution from
$W_{bk}(h_b)$ in $\overline {W}_{bk}$. We find
\begin{equation}
\frac{\epsilon}{4} \sum_b \sum_k m_b m_k \frac {v_{bk}^2}{\rho_b} \left (
\nabla_b W_{bk} ( \delta_{ab} - \delta_{ka} ) + \frac{\partial W_{bk}}{\partial h_b} H'_b
\frac {\partial \rho_b}{\partial {\bf r}_a} .
\right )
\end{equation}
with a similar term from the $W_{bk}(h_k)$ in $\overline {W}_{bk}$. Combining these terms
the contribution not involving $\partial W/\partial h$ is
\begin{equation}
    \frac{m_a \epsilon}{4} \sum_b m_b \frac{v_{ab}^2} {\overline {\rho}_{ab} } \left (
\nabla_a W_{ab}(h_a) + \nabla_a W_{ab}(h_b) \right ).
\end{equation}
The $\partial W/\partial h$ terms give the terms involving $H'$ in (4.9).

\section{ The continuum acceleration equation}
\setcounter{equation}{0}

From (4.9) the first term we consider is
\begin{equation}
 \frac {\epsilon}{2}\sum_b m_b   \frac {( {\bf \widehat v}_b - {\bf \widehat v}_a
)^2}{\overline {\rho}_{ab} }  \nabla_a W_{ab}. 
\end{equation}
To simplify the analysis we only consider the two dimensional case and assume the
density is constant. Without loss of generality we consider only the $x$ component.

The continuum limit of the summation is
\begin{equation}
\frac{\epsilon}{2} \int [{\bf v(r) - v(r^\prime}) ]^2 \frac {\partial W}{\partial
x^\prime}dx^\prime dy^\prime.
\end{equation}
Expanding ${\bf v}({\bf r}^\prime)$ about ${\bf r}$ to second order, and only retaining
terms of odd parity in $(x^\prime - x)$ and even parity in $(y^\prime -y)$, (B.2) becomes
\begin{equation}
\frac {\epsilon}{2}  \int \left ( - \Delta x^3 \frac {\partial^2 v^x}{\partial^2 x^2}
- \Delta x \Delta y^2 \frac {\partial^2 v^x}{\partial^2 y^2} - 2 \Delta x \Delta y^2
\frac{\partial v^x}{\partial y} \frac {\partial^2 v^x}{\partial x \partial y} + (v^x
\rightarrow v^y) \right ) \frac {\partial W}{\partial x^\prime} d x^\prime dy^\prime,
\end{equation}
where $v^x$ and $v^y$ denote the $x$ and $y$ components of ${\bf v}$ and $\Delta x$ and
$\Delta y$ denote $(x^\prime -x)$ and $( y^\prime - y)$ respectively. The notation $v^x
\rightarrow v^y$ means include a set of terms identical to the previous but with $v^x$
replaced by $v^y$. After integration by parts and noting that 
\begin{equation}
\int (x-x^\prime)^2 W dx^\prime dy^\prime = \int (y-y^\prime)^2 W dx^\prime
dy^\prime , 
\end{equation}
(B.2)  becomes
\begin{equation}
\alpha^2 \left ( 3 \frac {\partial v^x}{\partial x} \frac {\partial^2
v^x}{\partial x^2} + \frac {\partial v^x}{\partial x} \frac {\partial^2 v^x}{\partial 
y^2} + 2 \frac{\partial v^x}{\partial x} \frac {\partial^2 v^x}{\partial x \partial y}
+ (v^x \rightarrow v^y) \right).
\end{equation}
Combining terms we can write the previous expression as
\begin{equation}
\alpha^2 \left ( \frac {\partial v^x}{\partial x} \nabla^2 v^x + \frac
{\partial v^y}{\partial x} \nabla^2 v^y + \frac {\partial}{\partial x} ( \nabla v^x \cdot
\nabla v^x + \nabla v^y \cdot \nabla v^y) \right ).
\end{equation}
or, on replacing $v^x$ and $v^y$ by $v^1$ and $v^2$ respectively 
\begin{equation}
\alpha^2 \left ( \frac {\partial v^\ell}{\partial x} \nabla^2 v^\ell  + \frac
{\partial}{\partial x} ( \nabla v^\ell \cdot \nabla v^\ell  \right ).
\end{equation}
with summation on repeated indices.

The remaining contribution from (4.9) is 
\begin{equation}
-\frac{\epsilon}{4}\sum_b m_b \left ( \frac{\zeta_a}{\rho_a^2 \Omega_A } +
\frac{\zeta_b}{\rho_b^2
\Omega_b} \right ) \nabla_a W_{ab}(h).
\end{equation}
The continuum expression for $\zeta$ is 
\begin{equation}
\zeta = \int \rho({\bf r}') [ {\bf v}({\bf r}) - {\bf v} ({\bf r}')]^2 W({\bf r - r'})
{\bf dr}'.
\end{equation}
Expanding the velocity difference in a Taylor series as before, we get the approximation
\begin{equation}
\zeta = 2 \alpha^2 \nabla v^\ell \cdot \nabla v^\ell.
\end{equation}
Furthermore, noting $\rho$ is constant,
\begin{equation}
\sum_b m_b \frac{\zeta_a}{\rho^2} \nabla_a W_{ab} \rightarrow \frac{\zeta_a}{\rho} \int
\nabla_{{\bf r}} W {\bf dr}' .
\end{equation}
This integral vanishes because it has odd symmetry. The remaining term is
\begin{equation}
\sum_b m_b \frac{\zeta_b}{\rho^2} \nabla_a W_{ab} \rightarrow \frac {\nabla_a
\zeta_a}{\rho}.
\end{equation}
Multiplying (B.12) by $-\epsilon/4$ and combining with (12.7) gives 
\begin{equation}
\alpha^2 \left ( \frac {\partial v^\ell}{\partial x} \nabla^2 v^\ell  + \frac12 \frac
{\partial}{\partial x} ( \nabla v^\ell \cdot \nabla v^\ell  \right ).
\end{equation}
which, apart from the pressure term, gives the continuum form of the SPH force/mass.

The relevant term in Holm's equation (see (4.22)) is 
\begin{equation}
-v^\ell \frac {\partial \widehat v^\ell}{\partial x} +  \frac{\partial}{\partial x} 
\left ( \frac{1}{2} {\bf \widehat v}\cdot {\bf
\widehat v} + \frac {\alpha^2}{2} \frac {\partial \widehat v^j}{\partial x^\ell}\frac
{\partial \widehat v^j}{\partial x^\ell} \right ),
\end{equation}
which can be written 
\begin{equation}
(\widehat v^\ell -  v^\ell) \frac {\partial \widehat v^\ell}{\partial x} +  \frac
{\alpha^2}{2} \frac {\partial}{\partial x} \left (\frac {\partial \widehat v^j}{\partial
x^\ell}\frac {\partial \widehat v^j}{\partial x^\ell} \right ).
\end{equation}
The expression for the smoothed velocity can be inverted to give ${\bf v}$ in the form 
\begin{equation}
{\bf v} = {\bf \widehat v} -  \alpha^2 \nabla^2 {\bf \widehat v},
\end{equation}
when the density is constant. Substituting this expression into (B.15) we get
\begin{equation}
\alpha^2 \left ( \nabla^2  \widehat v^\ell \frac {\partial \widehat v^\ell}{\partial x}+ 
 \frac {1}{2} \frac {\partial}{\partial x} \left (\frac {\partial \widehat v^j}{\partial
x^\ell}\frac {\partial
\widehat v^j}{\partial x^\ell} \right) \right ). 
\end{equation}
Expression (B.13) the continuum limit of the SPH term and (B.17), the continuum term in
Holm's equation, are therefore identical.

\enddocument